\documentclass[aps,prb,superscriptaddress,showpacs, preprint]{revtex4}
\usepackage{graphics}
\usepackage{graphicx}
\usepackage{amsmath}
\usepackage{times}
\usepackage{bm}
\usepackage{amssymb}
\newcommand {\be} {\begin{equation}}
\newcommand {\bea} {\begin{eqnarray} \nonumber }
\newcommand {\ee} {\end{equation}}
\newcommand {\eea} {\end{eqnarray}}

\begin{document}
\title{Anisotropic profiles in the spin polarization of  multichannel semiconductor  rings
  with Rashba spin orbit coupling}
\author{L. E. Segura}
\affiliation{Centro At\'omico Bariloche and Instituto Balseiro, Comisi\'on
Nacional de Energ\'\i a At\'omica, 8400 San Carlos de Bariloche, Argentina.}
\author{M. J. S\'anchez}
\email{majo@cab.cnea.gov.ar}
\affiliation{Centro At\'omico Bariloche and Instituto  Balseiro, Comisi\'on
Nacional de Energ\'\i a At\'omica, 8400 San Carlos de Bariloche, Argentina.}

\begin{abstract}
We investigate the  spin accumulation effect  in eccentric  semiconductor multichannel rings with  Rashba spin-orbit interaction and  threaded by a magnetic flux.
Due to the finite eccentricity, the spin polarization  induced at the borders of the sample is anisotropic and
exhibits  different patterns and intensities at specific  angular directions. 
This  effect, reminiscent of   the   spin polarization drift  induced  by the application of an in plane  electric field,  could  be used to manipulate and functionalize the  spin polarization in electronic nanorings.

\end{abstract}
\maketitle

\section{Introduction}
The spin manipulation  and detection in {\em all electrical} devices have been
the subjects of an impressive amount of research  in the last years \cite{spinbook}.  
In particular semiconductors in  which the spin orbit (SO) interaction plays a significant role, are 
the main candidates to  accomplish such challenges.  Among them, ring shaped geometries are particularly attractive  to analize different interference phenomena in the presence of SO effects.
As an example, Aharonov Bohm (AB) rings with uniform SO interaction  have been  proposed  as spin interference devices \cite{nitta99} to explore the  non trivial spin dependent Aharonov-Casher (AC) phase \cite{ahar84,berry84,AronovG93}.   
Indeed, in recent years  the AC effect has been measured in a series of  transport  experiments on AB  conductance oscillations performed  in semiconductor rings, for different intensities of the SO interaction \cite{morpu98,meijer04,konig06,Bergsten06}. 

So far most of the  theoretical analysis of SO effects in quantum rings  have been  restricted to 1D geometries \cite{meir89,loss92,balatsky93,chaplik95,sgz02,lobo08}. 
However realistic rings  have a multichannel nature and many interesting phenomena  rely on this fact. 
As an example, in Ref.\onlinecite{loz05} it was shown that in a   AB multichannel ring with Rashba SO coupling \cite{rashba84}, a spin accumulation effect develops at the borders of the sample. Eventhough for an even number of electrons, a finite spin polarization in the direction perpendicular to the plane of the ring is generated and  can be controlled with the magnetic flux.  
This phenomena, although  sharing  some analogies with the intrinsic Spin Hall Effect (SHE) studied in bar or strip geometries designed on 2DEG \cite{kato04,Sinova04,wunder05},  does not require neither external  currents nor electric fields or voltage drops  applied \cite{usaj05,reynoso2}.

Actually in   real   semiconductor quantum rings (QRs)  imperfections of the structure often occurs. Micrograph views of GaAs QRs suggests that real rings may present some degree of eccentricity \cite{yau02}.
In addition, recent advances in oxidation lithography  enables the fabrication of semiconductor QRs  \cite{grbic07} in which non perfect symmetric structures can  be constructed.

The effect of a finite eccentricity has been considered in previous studies of energy spectrum and  electric polarization in QRs
\cite{egydio04,bruno05}. However these works have not taken into account the SO interaction, which is particularly strong in most of the QRs heterostructures.
 
The goal of the present work is to show that by a combined effect of  the Rashba SO  interaction and the finite eccentricity,  the spin accumulation  develops  anisotropic patterns along the sample. This effect, that shall be described in detail later on,  will  lead to different intensities  of the  total spin polarization in the angular direction. 
The typical spatial separation  of the anisotropic patterns may render possible   to   sense the effect employing usual magneto optical  detection techniques \cite{crooker05}.

The article is organized as follows. Sec.\ref{se2} introduces the model hamiltonian for the eccentric QR with Rashba SO interaction. After giving  some details of the calculations, the energy spectrum and eigenfunctions  are obtained for different magnetic fluxes. These results are analyzed and compared  to the ones of the  concentric QR.
In Sec.\ref{se3} the spin polarization is computed and the  different anisotropic patterns in the spin accumulation effect are analyzed in detailed.
The last section \ref{se4} is devoted to  the conclusions and to elaborate on the experimental feasibility to detect the anisotropic profiles in the spin accumulation effect.

\section{Eigenfunctions and eigenvalues of the eccentric QR with Rashba SO interaction}
\label{se2}
%Therefore, we believe that these structures tailor a  detailed investigation of  the electronic spectrum of eccentric QRs  
% both the ABE and the SO can play a relevant role. Moreover controlling the
 
\begin{figure}[t]
\includegraphics[height=6cm,clip]{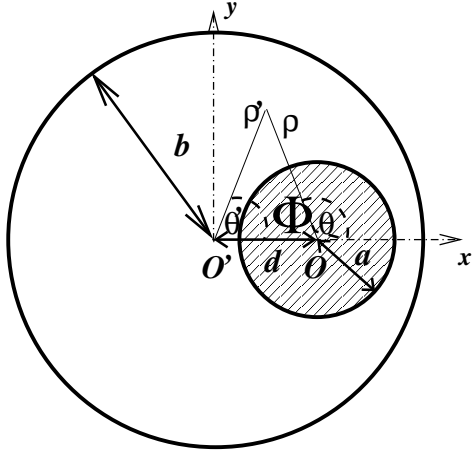}
\caption{Schematic of the  annular cavity with RSO threaded by a finite flux $\Phi$, where $a$ and $b$ are the internal and external radii respectively, and $d$ the eccentricity. $\rho (\rho'),\theta(\theta^{'})$ are the polar coordinates referred to the origin $O (O^{'})$.}
\label{fig1}
\end{figure}

We start by considering a 2D  electron gas in the $xy$ plane confined to an
 annular region delimited by two circles of radii $a$  and $b$  whose centers $O$ and $O^{'}$ are  separated  by a 
 distance $d < b -a $, that defines the eccentricity of the QR. A magnetic flux $\Phi$ threads the structure 
 (see Fig.\ref{fig1}). 
  
The single particle Hamiltonian describing an electron of effective
mass $m^{*}$ subjected to the Rashba spin orbit (RSO) coupling reads
\be
H=\frac{{\bf p}^2}{2m^{*}}+V+\frac{\alpha}{\hbar} ({\bf p} 
\times {\bf \hat{\sigma}}) \cdot \hat{z} \;,
\label{ham}
\ee
where $\alpha$ is the strength of the Rashba spin orbit (RSO) coupling 
 and the Pauli matrices ${\bf \hat{\sigma}}$ are defined as standard.
Let the polar coordinates be  $(\rho, \theta)$ and  $(\rho', \theta')$ with respect to $O$ and $O'$ respectively.
The confining potential defining the eccentric QR is 
\be \label{pot} 
V(\rho, \theta)= 
\left\{
\begin{array} {l} 
0   \; \; \mbox{for   $a< \rho < \rho_{ext} \equiv \sqrt{b^{2}+d^{2}-2 \; b\; d \; \cos\theta\sqrt{1-(\frac{d}{b}\sin\theta)^2}}$ } \\ 
\infty  \; \; \mbox{otherwise} \; .
\end{array} 
\right.  
\ee 
The vector potential 
which is introduced in the Hamiltonian via  the substitution, ${\bf p}= \hbar {\bf k}=-i \hbar{\bf \nabla} - 
\frac{e}{ c}{\bf A}$, is written in  the axial gauge as
${\bf A} = (\Phi / 2 \pi \rho) \hat {\theta}$ .
Using , 
$\hat{\sigma}_{\rho}= \cos\theta \hat{\sigma}_{x} + \sin\theta \hat{\sigma}_{y}$ and
$\hat{\sigma}_{\theta}= - \sin\theta \hat{\sigma}_{x} + \cos\theta \hat{\sigma}_{y}$ we 
can rewrite  the  Hamiltonian  as
\bea \label{h2d}
H=-\frac{\hbar^2}{2m^{*}}\left[ 
\frac{1}{\rho} \partial_{\rho}\left(\rho \partial_{\rho} \right)- 
\frac{1}{\rho^2}(i \partial_{\theta} + \nu)^2\right] \\  
+ i  \alpha \hat{\sigma}_{\theta} \partial_{\rho} - \frac{\alpha}{\rho} \hat{\sigma}_{\rho} 
\Bigl( i \partial_{\theta} + \nu \Bigr) \, \, ,
\eea
where  $\nu=\frac{\Phi}{\Phi_0}$ is the magnetic 
flux in units of the flux quantum $\Phi_0= {h c}{/ e} $.

In what follows it will be useful to define the dimensionless coordinates  $\xi=\rho/b$, $\xi' = \rho'/ b$, the aspect ratio $\lambda = b/a$ and 
\be
 \epsilon=\frac{2 m^{*} b^2}{\hbar^2} E \equiv \frac{E}{E_0} \,\,,\,\, \beta=2 \frac{m^{*} \alpha
   }{\hbar^2} b \; .
\ee

Due to the RSO, the bulk spectrum has two branches \cite{rashba84}
\be 
\epsilon= {(b k)}^2\pm \beta \;{(b k)}\;,
\ee
and for a given value of $\epsilon$ there are two non-trivial
solutions for the momentum $k$ that we denote $k^+$ and $k^-$ respectively. 

For the concentric ring ($d=0$),  the total angular momentum  $J_z=l_z+\frac{1}{2} \hbar \hat{\sigma}_z$ is a constant of motion and  $j=m+\frac{1}{2}$ is a  good quantum number even for finite RSO coupling. In Ref.\onlinecite{loz05} the complete set of  eigenfunctions (eigenspinors labeled by the quantum number $j=m+1/2$)  and  eigenvalues have been found  for the concentric QR with RSO. 

For a finite $d >0$, the angular momentum is  no longer conserved. In this case, we expand the solution of the Helmholtz equation associated to the hamiltonian Eq.(\ref{h2d})  in a basis of eigenfunctions of the total angular momentum, expressed in polar coordinates $(\rho,\theta)$ referred to the origin O (see Fig. \ref{fig1}), 
\be \label{eig}
\psi(\rho,\theta)=\left(
\begin{array}{c}
\sum_{m=-\infty}^{\infty}e^{im\theta}\varphi^{\uparrow}_{m-\nu}\left( \rho\right)  \\
 \\
\sum_{m=-\infty}^{\infty}e^{i(m+1)\theta}\varphi^{\downarrow}_{m-\nu+1} \left( \rho\right)\; ,
\end{array}
\right)
\ee
where  $\varphi^{\uparrow}_{m-\nu}\left( \rho\right) $ and $\varphi^{\downarrow}_{m-\nu+1}\left( \rho\right)$ are linear combinations of Bessels functions of first and second kind $J_{m-\nu}(k^{+,-}\rho), J_{m-\nu+1}(k^{+,-}\rho)$ and $Y_{m-\nu}(k^{+, -}\rho)$, $Y_{m-\nu+1}(k^{+, -}\rho)$ that satisfy  the boundary condition at the inner circle  $\psi (\rho=a,\theta)= 0 ; \forall \theta$, {\em i.e}:

\begin{eqnarray}
\varphi^{\uparrow}_{m-\nu}\left( \rho\right)&=&A_{m}\left(F'_{m-\nu}J_{m-\nu}(k^{+}\rho)+F_{m-\nu}Y_{m-\nu}(k^{+}\rho)+J_{m-\nu}(k^{-}\rho)\right)\\ \nonumber
&+&B_{m}\left(G'_{m-\nu}J_{m-\nu}(k^{+}\rho)+G_{m-\nu}Y_{m-\nu}(k^{+}\rho)+Y_{m-\nu}(k^{-}\rho)\right)\\ \nonumber
\varphi^{\downarrow}_{m-\nu+1}\left( \rho\right)&=&A_{m}\left(F'_{m-\nu}J_{m-\nu+1}(k^{+}\rho)+F_{m-\nu}Y_{m-\nu+1}(k^{+}\rho)-J_{m-\nu+1}(k^{-}\rho)\right) \\ \nonumber
&+&B_{m}\left(G'_{m-\nu}J_{m-\nu+1}(k^{+}\rho)+G_{m-\nu}Y_{m-\nu+1}(k^{+}\rho)-Y_{m-\nu+1}(k^{-}\rho)\right) \; .
\end{eqnarray}
with the  coefficients $F_{m-\nu},F'_{m-\nu},G_{m_-\nu}$ and $G'_{m-\nu}$ given by :
\begin{eqnarray}
 F_{m-\nu}&\equiv&D\left(J_{m-\nu+1}(k^{-}a)J_{m-\nu}(k^{+}a)-J_{m-\nu}(k^{-}a)J_{m-\nu+1}(k^{+}a)\right) \nonumber \\
 F'_{m-\nu}&\equiv&D\left(J_{m-\nu+1}(k^{-}a)Y_{m-\nu}(k^{+}a)-J_{m-\nu}(k^{-}a)Y_{m-\nu+1}(k^{+}a)\right) \nonumber \\
 G_{m-\nu}&\equiv&D\left(Y_{m-\nu+1}(k^{-}a)J_{m-\nu}(k^{+}a)-Y_{m-\nu}(k^{-}a)Y_{m-\nu+1}(k^{+}a)\right) \nonumber \\
 G'_{m-\nu}&\equiv&D\left(Y_{m-\nu+1}(k^{+}a)Y_{m-\nu}(k^{-}a)-Y_{m-\nu}(k^{+}a)Y_{m-\nu+1}(k^{-}a)\right) \nonumber \; ,\\
 D &=& {\left(J_{m-\nu}(k^{+}a)Y_{m-\nu+1}(k^{+}a)-J_{m-\nu+1}(k^{+}a)Y_{m-\nu}(k^{+}a) \right)}^{-1} \;.
\end{eqnarray}

The boundary condition $\psi (\rho_{ext}, \theta)=0 \; ,\forall \theta $  leads to a very complicated set of equation for determining the eigenvalues. One way to tackle this problem is to start by expanding the function  $\psi (\rho,\theta)$ into a Fourier series in $\theta^{'}$; 
\be
\psi(\rho,\theta)= \left(
\begin{array}{c}
\sum_{n= -\infty}^{\infty}\sum_{m=-\infty}^{\infty}Q_{mn}^{\uparrow}e^{in\theta'} \\ \nonumber
\sum_{n=-\infty}^{\infty}\sum_{m=-\infty}^{\infty}Q_{mn}^{\downarrow}e^{in\theta'}
\end{array}
\right)
\ee
where the matrices  $Q_{mn}^{\uparrow}$ and $Q_{mn}^{\downarrow}$ are defined as :
\begin{eqnarray}
\label{qmn}
& &Q_{mn}^{\uparrow}=\frac{1}{\pi}\int_{-\pi}^{\pi}\varphi_{m-\nu}^{\uparrow}\left( \rho\right) e^{im\theta}e^{-in\theta'}d\theta'\\ \nonumber
& &Q_{mn}^{\downarrow}=\frac{1}{\pi}\int_{-\pi}^{\pi}\varphi_{m-\nu+1}^{\downarrow}\left( \rho\right) e^{i(m+1)\theta}e^{-in\theta'}d\theta' \;.
\end{eqnarray}

The fully calculations of the above integrals  are rather cumbersome since  
$\rho$ and $\theta$ are functions of $ \rho',\theta '$. To evaluate Eqs.(\ref{qmn}) at the external boundary  $\rho=\rho_{ext}$, {\em i.e} $\rho'= b$, we employ addition theorems of Bessel functions \cite{sing84} and solve the  integrals analytically by means of  a perturbative expansion to order ${\cal O}(2)$ in the parameter $\nu \; \delta \lesssim 1 $ ($\delta \equiv d/b$).
After a lengthly derivation, we  arrive to the following  set  of equations :
\bea \label{det}
&& \sum_{n}\sum_{m} \hat{Q}_{mn}^{\uparrow}=0  \nonumber \\
&& \sum_{n}\sum_{m}\hat{Q}_{mn}^{\downarrow}=0 \;,
 \eea
where it is understood that $ -\infty < n, m < \infty$   being
\begin{eqnarray}\label{qmnb}
\hat{Q}_{mn}^{\uparrow}&=&A_{m}\left(F'_{m-\nu}Z_{mn}(k^{+},J)+F_{m-\nu}Z_{mn}(k^{+},Y)+Z_{mn}(k^{-},J)\right)\nonumber \\
&+&B_{m}\left(G'_{m-\nu}Z_{mn}(k^{+},J)+G_{m-\nu}Z_{mn}(k^{+},Y)+Z_{mn}(k^{-},Y)\right) \nonumber \\
\hat{Q}_{mn}^{\downarrow}&=&A_{m}\left(F'_{m-\nu}W_{mn}(k^{+},J)+F_{m-\nu}W_{mn}(k^{+},Y)-W_{mn}(k^{-},J)\right)\nonumber \\
&+&B_{m}\left(G'_{m-\nu}W_{mn}(k^{+},J)+W_{m-\nu}Z_{mn}(k^{+},Y)-W_{mn}(k^{-},Y)\right) \; ,
\end{eqnarray}
with the functions  $Z_{mn}(k^{*},C)$ and  $W_{mn}(k^{*},C)$  defined as:
\begin{eqnarray}\label{zw}
 Z_{mn}(k^{*},C)&=&a_{1}C_{n-\nu}(k^{*}b)J_{n-m}(k^{*}d)-a_{2}C_{n-\nu+1}(k^{*}b)J_{n-m+1}(k^{*}d)\nonumber \\
 &+&a_{2}C_{n-\nu-1}(k^{*}b)J_{n-m-1}(k^{*}d)+a_{3}C_{n-\nu+2}(k^{*}b)J_{n-m+2}(k^{*}d)\nonumber \\
&+&a_{4}C_{n-\nu-2}(k^{*}b)J_{n-m-2}(k^{*}d))\nonumber \\
W_{mn}(k^{*},C)&=&a_{1}C_{n-\nu}(k^{*}b)J_{n-m-1}(k^{*}d)-a_{2}C_{n-\nu+1}(k^{*}b)J_{n-m}(k^{*}d)\nonumber \\
 &+&a_{2}C_{n-\nu-1}(k^{*}b)J_{n-m-2}(k^{*}d)+a_{3}C_{n-\nu+2}(k^{*}b)J_{n-m+1}(k^{*}d)\nonumber \\
&+&a_{4}C_{n-\nu-2}(k^{*}b)J_{n-m-3}(k^{*}d)) \;.
\end{eqnarray}
In the last set of equation we denote  $k^{*}= k^{+}, k^{-}$ and  the coefficients $a_{i}, i=1,2,3,4$ are $a_{1}=2-\frac{(\nu\delta)^2}{2}$,  $a_{2}=\nu\delta$, $a_{3}=\frac{\nu\delta^{2}}{2}\left(\frac{\nu}{2}-1\right)$ and   $a_{4}=\frac{\nu\delta^{2}}{2}\left(\frac{\nu}{2}+1\right)$.
 $C\equiv J_{{}}, Y_{{ }}$  is  a Bessel function of first or second kind  whenever it corresponds.

For a given set  of parameters  $\beta$, $\lambda$ and $\delta$, we solve  Eqs.(\ref{det})-(\ref{zw}) for different values  of the flux $\nu$. 
The number of equations (defined by  the maximum values of $m,n$) is increased  until convergence in the solution is reached. 
The (dimensionless) energies $\epsilon(k^{+},k^{-})$ are found by the bisection method with a   precision  
$\sim 10 ^{-10}$. We denote by $\psi_{\epsilon_i}$ the associated eigenspinor built with the expansion coefficients $(A_{m}, B_{m})$ that solve Eq.(\ref{det}) for a given $\epsilon_i$. 

In order to fix numerical estimates for  the parameters we consider
characteristic values extracted from experiments. 
Rings with external radius $b \sim 400-500 nm$  and an aspect ratio  $\lambda \sim 2$ have been recently employed as devices \cite{meijer04}. Typical values for the Fermi wavelength are  $\lambda_F \sim 40-50 nm$ that give $k_F \sim 0.1 nm^{-1}$.
%For  $R=400nm$ one gets  a maximum value of the      
%(dimensionless) Fermi energy  $\epsilon_{F}= (R k_F)^2 \sim 1600$.
For an effective  mass $m^{*} \sim 0.042 m_e$, a Rashba coupling constant $\alpha= 8 meV\cdot nm$ and $b \sim 400 nm$ we obtain $\beta \sim 4$. These parameters characterize the sample {\bf S} studied in the present work.

In the case of concentric QRs ($d=0$), the total angular momentum $J$ is a constant of motion but the SO interaction breaks the degeneracy between
states differing in one unit of $j$. The degeneracy between states with opposite values of $j$ is removed by the finite magnetic flux $\nu$,  being the  charge persistent currents the signature of this broken time reversal   symmetry \cite{deblock02}. 
The energy spectrum exhibits a structure of level crossings  as a function of $\nu$, that are a signature of the Aharonov-Casher phase generated by the RSO coupling \cite{loz05}.

For $d>0$ the total angular momentum is no longer conserved due to  the lost  of  rotational symmetry induced by the finite eccentricity. As a consequence, the  quantum spectrum exhibits  avoided crossings  that are the   manifestation of the dynamical tunneling  between formerly degenerated states \cite{egydio04}.

\begin{figure}[t]
\includegraphics[height=6cm,clip]{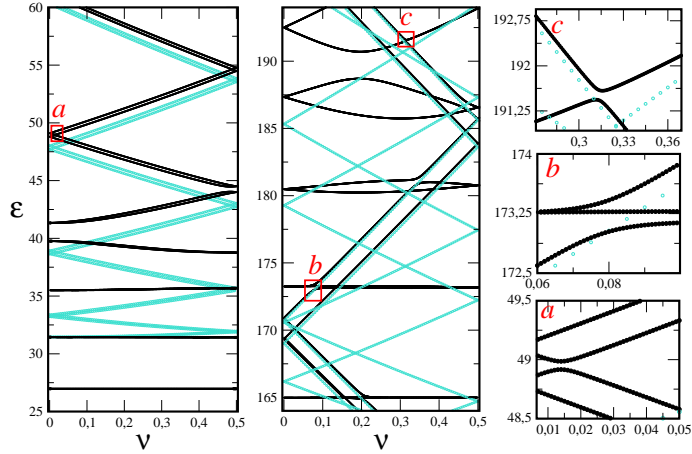}
\caption{Color on-line. Dimensionless energies $\epsilon$ as a function of the magnetic  flux 
$\nu$ for the  annular cavity $\bf{S}$ with RSO coupling defined in the text. The spectrum for the concentric ring $d=0$ is plotted by turquoise open circles  and for the eccentric QR, $d=0.05$, by  black  dotted lines. 
Left panel: Lowest eigenvalues. 
Central panel: Intermediate energy eigenvalues displaying avoided crossings for $d=0.05$. The regions inside the red boxes $a,b,c$ are enlarged in the right  panels to  show details of the  avoided crossings.}
\label{fig2}
\end{figure}

We display in Fig.\ref{fig2} the energy spectrum as a function of the flux $\nu$  for $d=0.05$ (black filled circles). This value of $d$ is
chosen to emphasize  that a rather small eccentricity  ($\sim 20 nm $ for the sample S) is enough to generate  qualitative changes in the quantum spectrum and its associated eigenfunctions, as we will show below. For  comparison we show the spectrum for  $d=0$ (turquoise empty circles), already studied in Ref.\onlinecite{loz05}. 

\begin{figure}[t]
\includegraphics[bb=50 50 770 302,width=20cm,clip]{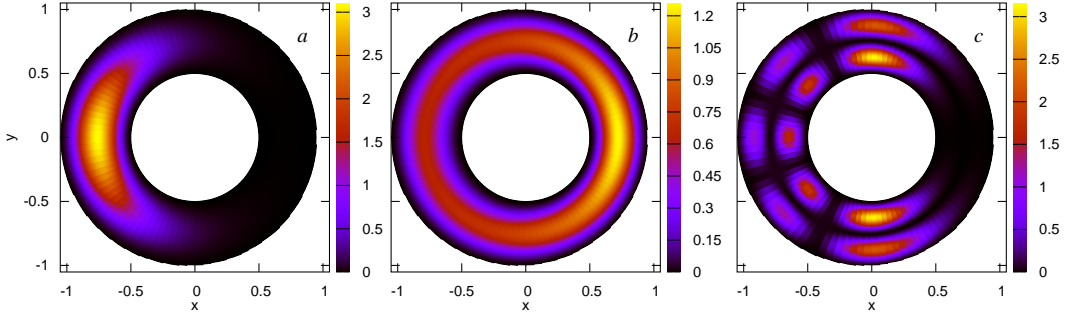}
\caption{Contour plots of the probability amplitudes in the sample $S$ for $\nu=0.2$ and an eccentricity $d=0.05$. 
a) ground state with $\epsilon= 26.96$, b) eigenstate with $\epsilon = 49.11 $ and c) eigenstate with  $\epsilon =164.97$. The values of $\epsilon$ are given at $\nu =0$.}
\label{fig3}
\end{figure}

The left panel of Fig.\ref{fig2} shows the lowest eigenvalues of the QR {\bf{S}}. Only the first transverse mode  is  active in this region and the spectrum shares some characteristics with the one  of a $1D$ ring (due to the symmetry respect to $\nu=0.5$ the spectrum is shown for $0 \le \nu \le 0.5$).
However, for $d=0.05$ the lower levels  are almost  flat   as a function of the flux, with associated 
 eigenfunctions that are highly localized in one side of the QR.  They are formed  mainly by contributions from 
 low angular momentum eigenstates of the concentric ring, being these states highly affected by the finite eccentricity.  
In Fig.\ref{fig3} a) we plot a contour plot of the  probability amplitude of  the ground
state $|\psi_{\epsilon_{1}}|^{2}$ at $\nu=0.2$, which it is mostly concentrated around  $\theta= \pi$.
Notice that the  finite  eccentricity produces  a similar effect than an in plane electric field, although  it does not depend on the charge of the particle\cite{bruno05}.
 
The other type of levels which lay higher in the spectrum have  a finite slope $\partial \epsilon_{i} /\partial \nu$ (current carrying states) and  resemble  the ones of  the unperturbed ring   with higher  values of the total angular momentum $j$. However, due to the finite eccentricity, energy splittings  and avoided crossings  appear at $\nu \geq 0$. A blow up image of one of these avoided crossings is displayed in Fig.\ref{fig2} in the box label a). Regarding  the associated eigenfunctions, they   are mainly generated by combination of  whispering gallery modes of the concentric ring \cite{egydio04} and therefore are not much affected by the finite eccentricity.  In  Fig.\ref{fig3} b) we display for $\nu = 0.2$ a contour plot of the  probability amplitude for one of these states ($\epsilon= 49.11$)  which  is  quite  homogeneously distributed in the angular direction. 

When higher  transverse channels are  active, the spectrum for the eccentric QR is more involved. It displays  additional avoided crossings generated by the mixing of levels that for $d=0$ were degenerated belonging to  different transverse channels. Two detailed images of these avoided crossings are  displayed in the right panel of Fig.\ref{fig2} in boxes $b$ and $c$.

Again, localized (flat levels) and current carrying states coexist in this region of the spectrum. As an example, in  Fig.\ref{fig3} c) we plot for $\nu = 0.2$ a contour plot of the  probability amplitude for one of the angular localized  states with $\epsilon= 164.97$. In displays  maxima and minima along the radial direction  which are the fingerprint of  basis states belonging  to the second transverse channel of the concentric QR.   

Taking into account that a given eigenspinor $\psi_{\epsilon_{i}}$ is an expansion in a basis of functions of total angular momentum, its   probability amplitude has  an angular  profile which  originates in the  interference  terms $\propto \sum_{m,m'} \left(\varphi^{\uparrow}_{m-\nu}\varphi^{\uparrow}_{m'-\nu}+ {\varphi^{\downarrow}_{m-\nu+1}}\varphi^{\downarrow}_{m'-\nu+1}\right) \cos (m-m')\theta$ (see Eq.\ref{eig}). 
As we have shown  in Fig.\ref{fig3}, the  angular profile  depends on the specific eigenstate.
This effect will  also influence the  spin accumulation describe in the next section.

\section{Spin accumulation effect}\label{se3}

As a result of the RSO interaction, the spin projection is not a good quantum number. Given an eigenspinor $\psi_{\epsilon_{i}}$ whose general form is given by the Eq.\ref{eig} the mean value  of the z-projection of the spin density  is proportional to
\bea \label{dsigma}
&&\sigma_z(\xi,\theta)=\psi_{\epsilon_i}^{\dagger}\hat{\sigma}_z \psi_{\epsilon_i} = 
\sum_{m=-M}^{M}\left( |\varphi^{\uparrow}_{m-\nu}|^2 - |\varphi^{\downarrow}_{m-\nu+1}|^2 \right) + \nonumber \\
 &&+ 2\sum_{m=-M}^{M}\sum_{m'=m+1}^{M}\left(\varphi^{\uparrow}_{m-\nu}\varphi^{\uparrow}_{m'-\nu}-
 {\varphi^{\downarrow}_{m-\nu+1}}\varphi^{\downarrow}_{m'-\nu+1}\right) cos{((m-m')\theta)} \;,
\eea
where we have used the fact that the radial part of the up and down  components of an eigenspinor  are real functions. Thus the  products  ${\varphi^{\uparrow,\downarrow}_{m-\nu}}\varphi^{\uparrow,\downarrow}_{m'-\nu}$ are even under the interchange 
$ m \leftrightarrow m'$. The  sum is for $ -M \leq m \leq M$, where $M$ is given by the cut  off in the expansion in the basis of total angular momentum eigenspinors.

In Ref.\onlinecite{loz05} it has been shown that in a  multichannel concentric QR with RSO interaction, a spin accumulation effect (SAE) develops for finite values of the magnetic flux $\nu$. The SAE is the tendency of the total spin density for N particles,  ${\Sigma}_z \equiv \sum_N \sigma_z$, to be different from zero,  positive on one border of the sample and  negative  on  the other one \cite{usaj05}. 

In the concentric geometry and for   $\nu=0$, states with opposite value of $j$ have   opposite values  of ${\sigma}_z \neq 0$ and then for even number of particles $N=2p$, is  ${\Sigma}_z = 0$. 
A finite $\nu \neq 0$ breaks the spatial symmetry between single particle states
with opposite value of $j$. This can be understood in terms of   the effective orbital index of the Bessel functions $m_{eff} \equiv m-\nu$, which turns to be
$m-\nu$ and $-m-1-\nu$ for $j_{1}=m+1/2$ and $j_{2}=-j_{1}$, respectively. Thus the modulus of $m_{eff}$  decreases  for $j_{1}$ and increases for  $j_{2}$. As a consequence  $\sigma_z $ is  ``pushed" toward the internal (external) boundary of the sample for the state label by $j_{1}$ ($j_{2}$). This is the main ingredient that contributes  to generate the SAE in the concentric QRs \cite{loz05},{\em i.e} a  total spin polarization $\Sigma_z$ that has opposite signs on each border of the sample $S$.
However neither $\sigma_z$ nor $\Sigma_z$ depend on the angle $\theta$.

\begin{figure}[t]
\includegraphics[bb=50 50 410 302,clip]{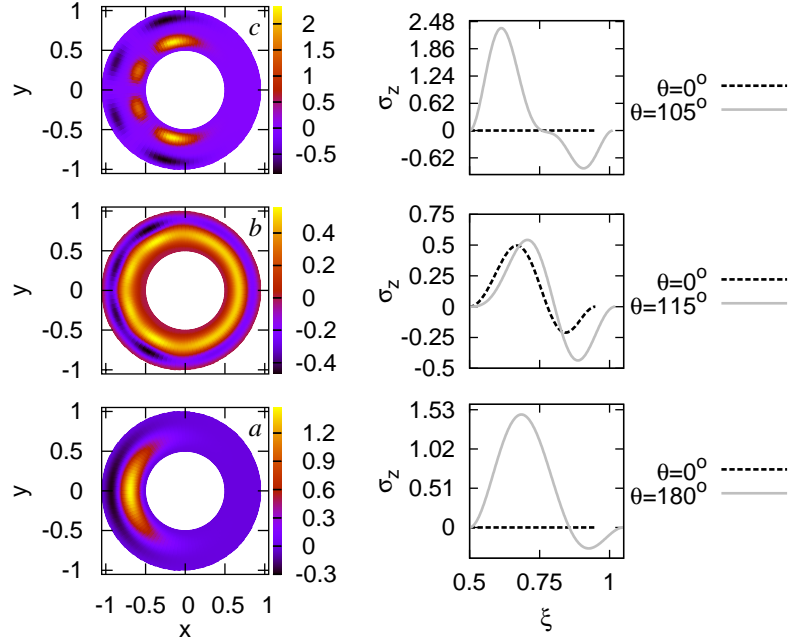}
\caption{Left panel:contour plot of the spin density  $\sigma_z(\xi,\theta)$ for the three eigenspinors displayed in
Fig.\ref{fig3}. The anisotropic pattern in  the angular direction is  clearly observed for a) and c), where for panel b) the angular profile  is almost uniform.  The right panels show $\sigma_z$ vs $\xi$ for two specific  angular directions selected in order to stress the effect. In all the cases is $\nu= 0.2$.}
\label{fig4}
\end{figure}

In the eccentric QR, the SAE develops an anisotropic angular profile due to the interference (last) term 
in Eq.\ref{dsigma}. The  particular profile  depends on   the  specific  eigenspinor, exhibiting  different characteristics for localized or current carrying states.  
As an example in  Fig.\ref{fig4}, we plot the contours plots of  $\sigma_z(\xi, \theta)$ calculated for the  three  eigenspinors analyzed in Sec.\ref{se2}. The anisotropic angular pattern  is  clearly observed for  the two localized states a) and c), where it is almost uniform for the current carrying state b).  

As the particle number  $N$ is increased to relevant experimental values, transverse channels are activated and the description of the effect becomes more involved. Nevertheless, mainly due to the existence of localized states  along all energy ranges,  the anisotropy in the SAE is a generic feature of the eccentric QR pierced by a  magnetic flux. 

\begin{figure}[t]
\includegraphics[bb=50 50 482 302,clip]{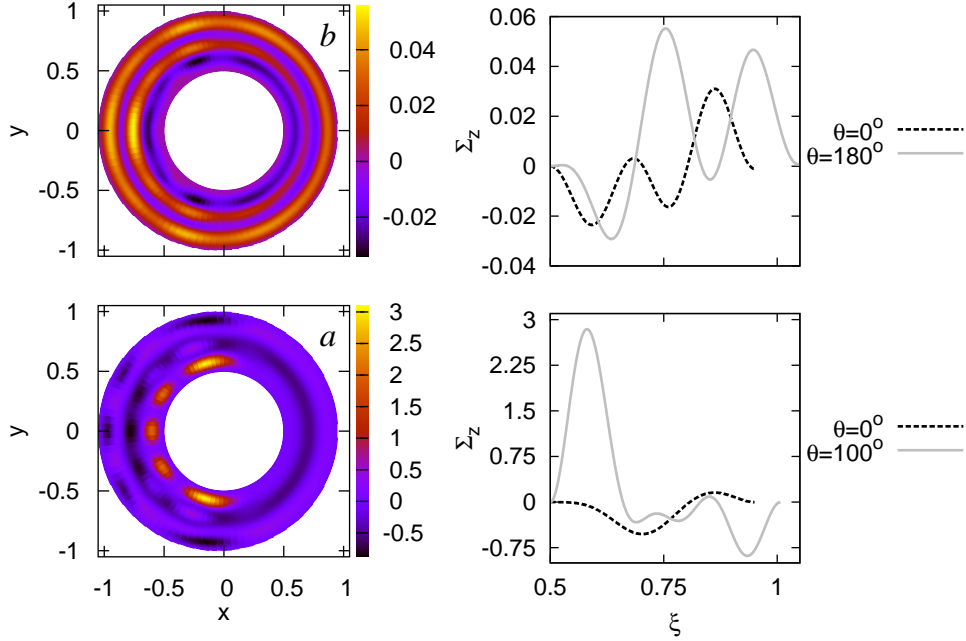}
\caption{Left panels: Contour plot of the  total spin density $\Sigma_z (\xi,\theta)$ for $N=100$ particles. Right panels: $\Sigma_z$ vs $\xi$ for the  two  angular directions $\theta$ selected to stress the anistropy. 
Lower (upper) panels correspond to  $\nu=0.2$ ($\nu=0.45$).}
\label{fig5}
\end{figure}

As an illustration, in the left panels of Fig.\ref{fig5} we show the contour plots of the total  spin density $\Sigma_z (\xi,\theta) \equiv \sum^{N} \sigma_z $   for  $N=100$  particles and  for two different values of the magnetic flux $ \nu =0.2, 0.45$.

The sensitivity of the SAE to the value of the magnetic flux $\nu$, is due  to the qualitative changes that the eigenspinors experience at the  avoided crossings in the flux-energy landscape. For this particular filling, the anisotropy in the SAE is  more pronounced for  $ \nu =0.2$ than for $ \nu= 0.45$. 
 
Besides the accumulation effect, the total spin density  per unit area $M_{z}$, is different from zero  and  its value 
depends  on   $\theta$. 
Let the differential  area be  $d A= (b^{2} -a^{2}) d\theta$ ($A= 0.75 \pi b^{2}$ is the area of sample $S$). It is easy to verify that $M_{z} (\theta) =\frac{4}{3}  \int_{.5}^{1} \Sigma_{z} (\xi,\theta) \; \xi \; d\xi $.  
 Taking  $ b= 0.4 \mu m$ and for  $N=100$ particles,  $M_z$ ranges from   $M_{z}(\theta=0)= -0.3 {\mu m}^{-2}$ to
 $M_{z}(\theta=100^{\circ})=0.48 {\mu m}^{-2}$ for  $\nu=0.2$. For  $\nu=0.45$, $M_{z}$ is generally smaller, but 
  the differences  are still important and maxima between  $M_{z} (\theta=0)= 1.33\times 10^{-3} {\mu m}^{-2}$ and 
  $M_{z}(\theta=180^{\circ})=-0.046 {\mu m}^{-2}$.
Thus, $M_z$ experiences important changes and even reverses  it sign  on  distances of the order $1 \mu m$.
This  is scketched in  the right panels of Fig.\ref{fig5}, where we show $\Sigma_z$ vs $\xi$ for  the two  angular directions $\theta$ especially  selected to stress the anistropy.

\section{Discussion and  Summary} \label{se4}

We have shown that a finite magnetic flux  in an eccentric  multichannel QR  with SO interaction induces a spin accumulation effect (SAE) that exhibits  an anisotropic profile in the angular direction.
The effect, which is reminiscent  to the  drift of the spin polarization  induced  by the application of an in plane  electric field, remains appreciable  for  filling numbers $(N\sim 100)$ which are experimentally realized in actual semiconductor QRs.

The spatial structure of the SAE which can be $\sim 1 \mu m$ in the sample $S$ (see Fig.\ref{fig5}), is not far from the sensitivity of detection methods based on scanning Kerr microscopy, which has been 
recently employed  to image spin polarization in semiconductor channels \cite{kato04,crooker05}.

In addition, the total spin density per unit area  $M_{z}$, reaches at some angular directions, values $\sim 0.5 {\mu m}^{-2}$.
These values  are comparable to the  spin densities  measured so far in the Spin Hall Effect in semiconductor channels \cite{kato04}.

In the presence of an external magnetic field perturbation, the spin magnetization should be proportional to  $M_z $. With the help of new experimental techniques  based on microresonators \cite{deblock02} or scanning magnetic force microscopes  that can reach resolutions of $\sim 10 nm $ \cite{magn08}, it should be feasible to  sense  differences in the values  of $M_z$ bewteen
sample regions separated $~1\mu m$.

We believe that our results could help to manipulate and functionalize the  spin polarization  in QRs with SO interaction, without using  applied electric fields or external currents.

\vspace*{2cm}

Partial financial support by ANPCyT Grants 06-483/03-13829 and  CONICET
are gratefully acknowledged. We would like to thank G. Usaj and C. Balseiro for helpful discussions.

\end{document}